\documentclass[10pt,aps,prd,floatfix,nofootinbib,superscriptaddress,preprintnumbers]{revtex4-2}
\pdfoutput=1

\usepackage{graphicx} 
\usepackage{amsmath,amssymb,bm}
\usepackage{xcolor}
\usepackage{mathtools}
\usepackage[colorlinks,allcolors=blue]{hyperref} 

\begin{document}

\preprint{TU-1298}

\title{Microlensing events and primordial black holes in the axionlike curvaton model}

\author{Kentaro Kasai}
\email{kkasai@icrr.u-tokyo.ac.jp}
\affiliation{ICRR, University of Tokyo, Kashiwa 277-8582, Japan}
\author{Masahiro Kawasaki}
\email{kawasaki@icrr.u-tokyo.ac.jp}
\affiliation{ICRR, University of Tokyo, Kashiwa 277-8582, Japan}
\affiliation{Kavli IPMU (WPI), UTIAS, University of Tokyo, Kashiwa, 277-8583, Japan}
\author{Kai Murai}
\email{kai.murai.e2@tohoku.ac.jp}
\affiliation{Department of Physics, Tohoku University, Sendai 980-8578, Japan}
\author{Shunsuke Neda}
\email{neda@icrr.u-tokyo.ac.jp}
\affiliation{ICRR, University of Tokyo, Kashiwa 277-8582, Japan}

\begin{abstract}
Recently, Subaru Hyper Suprime-Cam (HSC) observations found 12 candidates for microlensing events.
These events can be explained by primordial black holes (PBHs) with masses of $10^{-7}$--$10^{-6} M_\odot$ and a fraction of all dark matter of $f_\mathrm{PBH} = \mathcal{O}(10^{-1})$.
In this paper, we consider the PBH production in two types of the axionlike curvaton models, which predict an enhancement of the curvature perturbations on small scales.
We show that the microlensing events can be explained in the axionlike curvaton model and discuss the cosmological implications such as gravitational waves.
\end{abstract}

\maketitle

\section{Introduction}

Recently, Sugiyama \textit{et al.}~\cite{Sugiyama:2026kpv} reported that the Subaru Hyper Suprime-Cam (HSC) observations found 12 candidates (including 4 secure candidates) for microlensing events that could be caused by primordial black holes (PBHs)~\cite{Zeldovich:1967lct,Hawking:1971ei,Carr:1974nx,Carr:1975qj}.
PBHs are black holes formed by the gravitational collapse of large density fluctuations in the early universe.
They are also motivated by various cosmological considerations such as dark matter, gravitational wave events, and the origin of supermassive black holes.
The microlensing signal can be explained by PBHs with masses of $M_\mathrm{PBH} = 10^{-7}$\,--\,$10^{-6} M_\odot$ and density fraction to all dark matter of $f_\mathrm{PBH} = \mathcal{O}(10^{-1})$.

The mass distribution of PBHs is determined by the properties of density fluctuations, such as their amplitude, scale, and statistics.
To generate PBHs in a certain mass range, the power spectrum of the curvature perturbations should be enhanced on the corresponding scale.
In this work, we focus on the axionlike curvaton models.
The curvaton is a field responsible for generating primordial curvature perturbations instead of the inflaton~\cite{Enqvist:2001zp,Lyth:2001nq,Moroi:2001ct}.
In the axionlike curvaton models, an axion, a pseudo Nambu-Goldstone boson arising as a phase component of a complex scalar field spontaneously breaking a global U(1) symmetry, can play the role of the curvaton.
In this case, due to the inflationary evolution of the complex scalar field, the fluctuations of the curvaton field have a distinctive scale dependence~\cite{Kasuya:2009up}, which is discussed in the contexts of PBH formation~\cite{Kawasaki:2012wr,Bugaev:2013vba,Ando:2017veq,Ando:2018nge,Inomata:2020xad,Kawasaki:2021ycf,Ferrante:2023bgz,Inomata:2023drn} and the generation of gravitational waves~\cite{Kawasaki:2013xsa,Kawasaki:2021ycf,Inomata:2023drn}.

One of the remarkable properties of the axionlike curvaton models is the non-Gaussianity of the curvature perturbations~\cite{Sasaki:2006kq,Kawasaki:2011pd}, which is in contrast to the single-field slow-roll inflationary models.
The non-Gaussianity of curvature perturbations significantly affects PBH formation~\cite{Bullock:1996at,PinaAvelino:2005rm,Hidalgo:2007vk,Byrnes:2012yx,Young:2013oia,Young:2014oea,Young:2015kda,Young:2015cyn,Nakama:2016gzw,Pattison:2017mbe,Garcia-Bellido:2017aan,Atal:2018neu,Cai:2018dkf,Franciolini:2018vbk,Passaglia:2018ixg,Atal:2019cdz,Yoo:2019pma,Taoso:2021uvl,Riccardi:2021rlf,Escriva:2022pnz,Matsubara:2022nbr,Ferrante:2022mui,Yoo:2022mzl,vanLaak:2023ppj,Pi:2024lsu,Inui:2024fgk,Shimada:2024eec}.
The non-Gaussianity in the axionlike curvaton model can be parameterized by the local type non-Gaussianity, $f_\mathrm{NL}$, which can be positive or negative depending on the energy ratio of the curvaton field to radiation at the curvaton decay, $r_\mathrm{dec}$.
If $r_\mathrm{dec}\gg1$, the non-Gaussianity becomes negative, i.e., $f_\mathrm{NL} < 0$, and the PBH formation is suppressed.
Conversely, $r_\mathrm{dec} \lesssim 1$ predicts positive $f_\mathrm{NL}$, enhancing the PBH formation, while the curvaton contribution to the curvature perturbation is suppressed.
Anyway, the enhanced curvature perturbation predicted in the axionlike curvaton models can predict the PBH formation.

The enhanced curvature perturbation also sources primordial gravitational waves via the second-order effect~\cite{Tomita:1967wkp,Matarrese:1992rp,Matarrese:1993zf,Matarrese:1997ay,Noh:2004bc,Carbone:2004iv,Ananda:2006af,Baumann:2007zm,Saito:2008jc,Bugaev:2009zh,Saito:2009jt,Kohri:2018awv}, which are called scalar-induced gravitational waves (SIGWs) (see also Ref.~\cite{Domenech:2021ztg} for a review).
Thus, depending on the non-Gaussianity of the curvature perturbations, this scenario for the PBH formation can be limited by the overproduction of SIGWs or can be probed via gravitational wave observations.

There are two types of axionlike curvaton models: type I~\cite{Kawasaki:2012wr,Kawasaki:2013xsa,Bugaev:2013vba,Ando:2017veq,Inomata:2020xad,Ferrante:2023bgz} and type II~\cite{Ando:2018nge,Kawasaki:2021ycf,Inomata:2023drn}.
While an axionlike field works as a curvaton and enhances curvature perturbations in both types of models, the evolution of the complex scalar field is different.
In the type I models, the radial component of the scalar field rolls down the potential from a large value, which results in a blue-tilted spectrum.
In the type II models, the radial component is stabilized at a small value in the early stage of inflation and rolls down the potential toward a larger field value in the later stage. 
Due to the evolution of the radial component and the effective mass of the phase component, the curvature perturbation has a peak on a certain scale.

In this paper, we investigate the PBH formation in type I and type II axionlike curvaton models motivated by the reported microlensing events.
We aim to assess whether the PBH abundance suggested by the microlensing events can be reproduced in each type.
Further, by studying the two types in parallel, we identify features that are generic or model-type dependent and highlight the corresponding imprints in the PBH mass function and the SIGW spectrum.
In particular, we consider three fiducial cases for the microlensing events: $r_\mathrm{dec} = 0.3$ and $3$ in the type II model and $r_\mathrm{dec} = 0.3$ in the type I model.
Due to the effects of non-Gaussianities, the curvature perturbations required to account for the microlensing events are larger for $r_\mathrm{dec} = 3$ than for $r_\mathrm{dec}  = 0.3$.
In addition, the PBH mass distribution becomes sharper for $r_\mathrm{dec} = 3$ than for $r_\mathrm{dec} = 0.3$.
When comparing type I and II models for $r_\mathrm{dec} = 0.3$, the peak amplitude of the curvature perturbations is similar while the spectral shape is distinct, which is inherited by the power spectrum of SIGWs.
In summary, all three cases can account for the PBH abundance expected from the microlensing events.
While $r_\mathrm{dec}$ or non-Gaussianities affect the shape of the PBH mass distributions and the amplitude of SIGWs, the difference between the type I and II models can be seen in the spectral shape of the SIGWs.
Moreover, we also consider a different parameter set in the type II model with $r_\mathrm{dec} = 0.3$ that accounts for all dark matter by PBHs with smaller masses, $M = \mathcal{O}(10^{-12}) M_\odot$.

Note that in the case of $r_\mathrm{dec} > 1$ in the type I model, the enhanced curvature perturbations reenter the horizon while the oscillating curvaton dominates the universe.
Due to the absence of radiation pressure, PBH formation is significantly enhanced during the matter-dominated era and requires a distinct treatment for evaluation, which is beyond the scope of this work (see also the discussion in Ref.~\cite{Inomata:2023drn}).

The rest of this paper is organized as follows.
In Sec.~\ref{sec: type II}, we briefly explain the setup of the type II axionlike curvaton model and evaluate the power spectrum of the curvature perturbations.
Then, we do the same for the type I axionlike curvaton model in Sec.~\ref{sec: type I}.
In Sec.~\ref{sec: PBH}, we discuss the PBH formation from non-Gaussian curvature perturbations and derive the PBH mass function in our scenario.
In Sec.~\ref{sec: GW}, we investigate the viability and observability of the SIGWs.
Sec.~\ref{sec: summary} is devoted to the summary of our results.

\section{Type II axionlike curvaton model}
\label{sec: type II}

First, we explain the setup of the type II axionlike curvaton model~\cite{Ando:2018nge,Kawasaki:2021ycf,Inomata:2023drn}.
In this scenario, we consider a real scalar field $I$, which plays the role of the inflaton, and a complex scalar field $\Phi$, whose phase component works as a curvaton field.
The Lagrangian is given by 
\begin{align}
    \mathcal{L}
    =
    \frac{1}{2} \partial_\mu I \partial^\mu I 
    + \partial_\mu \Phi \partial^\mu \Phi^*
    - V_I(I) - V_\Phi(I, \Phi)
    \ ,
\end{align}
where $V_I$ is the inflaton potential, and $V_\Phi$ is given by 
\begin{align}
    V_\Phi(I, \Phi)
    =
    \frac{\lambda}{4} \left( |\Phi|^2 - \frac{v^2}{2} \right)^2
    + g I^2 |\Phi|^2
    - v^3 \epsilon \left( \Phi + \Phi^* \right)
    \ .
\end{align}
Here, $\lambda$ and $g$ are dimensionless coupling constants, $v/\sqrt{2}$ is the vacuum expectation value (VEV) of $\Phi$, and $\epsilon \ll 1$ is a small constant explicitly breaking the U(1) symmetry under which $\Phi$ is charged.

\begin{figure}[t]
    \centering
    \includegraphics[width=.6\textwidth]{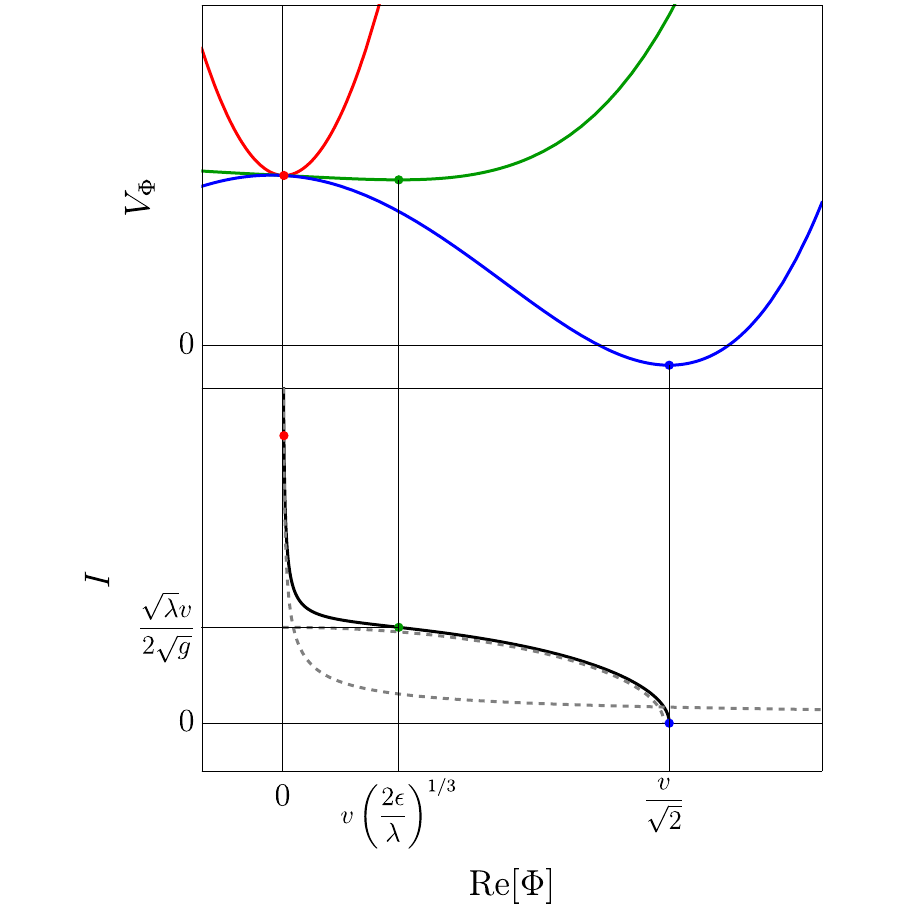}
    \caption{%
        Evolution of $V_\Phi$ and its minimum during inflation.
        The horizontal axis is the real part of $\Phi$.
        Upper panel: 
        The potential shape depending on $I$.
        $V_\Phi$ evolves in the order of red, green, and blue as $I$ decreases.
        The colored dots denote the potential minima of the potential with the corresponding colors.
        Lower panel: 
        The trajectory of the potential minimum.
        The colored dots are the same as in the upper panel.
        The gray dashed lines denote the approximated relations, $\Phi = \epsilon v^3/(gI^2)$ and $\Phi = \sqrt{v^2/2 - 2gI^2/\lambda}$, which applies for $I \gtrsim \sqrt{\lambda} v/(2 \sqrt{g})$ and $I \lesssim \sqrt{\lambda} v/(2 \sqrt{g})$, respectively.
    }
    \label{fig: V and I}
\end{figure}

In this scenario, we assume that $I$ decreases from a large positive value toward zero during inflation, while we do not specify the exact form of $V_I$ so far. 
Then, $\Phi$ is stabilized around the origin in the early stage of inflation.
For $I \gtrsim \sqrt{\lambda} v/(2 \sqrt{g})$, $\Phi$ is given by $\Phi \simeq \epsilon v^3/(gI^2)$ due to a balance between the second and third terms of $V_\Phi$.
When $t = t_\mathrm{peak}$ defined by $I = \sqrt{\lambda} v/(2 \sqrt{g})$, the mass term in $V_\Phi$ vanishes, and $\Phi$ is given by $\Phi = (2\epsilon/\lambda)^{1/3}v$.
For $I \lesssim \sqrt{\lambda} v/(2 \sqrt{g})$,
$\Phi$ moves towards $v/\sqrt{2}$ as $\Phi = \sqrt{v^2/2 - 2gI^2/\lambda}$ due to a balance between the first and second terms of $V_\Phi$.
We show how $V_\Phi$ and $\Phi$ evolve during inflation in Fig.~\ref{fig: V and I}.
Note that $\Phi$ evolves in the positive direction of the real axis due to the $\epsilon$ term.
As a result, cosmic strings are not formed in this scenario.

For later convenience, we decompose $\Phi$ into the background field and perturbations in the radial and phase directions as
\begin{align}
    \Phi 
    = 
    \frac{1}{\sqrt{2}} (\varphi_0 + \delta \varphi) e^{i \delta \sigma/\varphi_0}
    \ ,
    \label{eq: Phi decomposition}
\end{align}
where $\varphi_0$ denotes the background field of $\Phi$, and $\delta \varphi$ and $\delta \sigma$ denote perturbations in the radial and phase directions, respectively.
The effective masses of $\delta \varphi$ and $\delta \sigma$ depend on $\varphi_0$ as
\begin{align}
    m_{\delta \varphi}^2 
    &=
    \left. 
        \frac{\partial V_\Phi}{\partial \delta \varphi^2}
    \right|_{\delta \varphi = 0}
    =
    \frac{\lambda}{4} \left( 3 \varphi_0^2 - v^2 \right)
    + g I^2
    \ ,
    \\
    m_{\delta \sigma}^2 
    &=
    \left. 
        \frac{\partial V_\Phi}{\partial \delta \sigma^2}
    \right|_{\delta \sigma = 0}
    =
    \sqrt{2} \epsilon \frac{v^3}{\varphi_0}
    \ .
\end{align}
In the following, we assume $\lambda v^2 \gg H^2$ during inflation so that $\varphi_0$ rapidly rolls down to $\sim v$ after the destabilization.

Due to the evolution of $\varphi_0$, the power spectrum of $\delta \sigma$ has a significant scale dependence.
Let us define the time when the wavenumber $k$ exits the horizon during inflation by $t_k$.
Then, fluctuations of $\Phi$ at the horizon exit have the power spectrum of
\begin{align}
    \label{eq:power_spec_Hcross}
    \mathcal{P}_{\delta \varphi}(k, t_k)
    =
    \mathcal{P}_{\delta \sigma}(k, t_k)
    \simeq
    \left(\frac{H}{2\pi} \right)^2
    \ ,
\end{align}
where we assumed that the Hubble parameter $H$ is constant during inflation.
Here, the power spectrum is defined as $\langle \delta\varphi(\bm{p},t_p) \delta\varphi(\bm{q},t_q) \rangle = (2\pi)^3 \delta^{(3)}(\bm{p}+\bm{q}) P_{\delta \varphi}(p,t_p)$ with $\mathcal{P}_{\delta \varphi}(p,t_p) = p^3 P_{\delta \varphi}(p,t_p)/(2\pi^2)$.%
\footnote{
    Precisely speaking, when the effective mass is much larger than the Hubble parameter ($m_{\delta\sigma},\, m_{\delta\varphi} \gg H$), the fluctuations at the horizon crossing are more suppressed than those given by Eq.~\eqref{eq:power_spec_Hcross}.
    However, around the peak of the power spectrum where we are interested, the effective mass is not much larger than the Hubble parameter ($m_{\delta\sigma}/H\lesssim \mathcal{O}(1)$), and  Eq.~\eqref{eq:power_spec_Hcross} is approximately satisfied.
}

After the horizon exit, the fluctuations are suppressed if its effective mass is greater than $H$.
In addition, those of $\delta \sigma$ also evolve due to the time evolution of $\varphi_0$.
When we disregard the effect of the effective mass, $\mathcal{P}_{\delta \sigma}$ evolves so that $\mathcal{P}_{\delta \theta}$ becomes constant with $\delta \theta \equiv \delta \sigma/\varphi_0$.
Thus, to see the effect of the effective mass, it is convenient to see the evolutions of $\delta \theta$ and $\mathcal{P}_{\delta \theta}$.
Due to the effective mass of $\delta \sigma$, $\delta \theta$ is suppressed after the horizon exit as
\begin{align}
    \frac{\delta \theta(t_\mathrm{end})}{\delta \theta(t_k)}
    =
    \exp \left(
        - \frac{3}{2} \int_{t_k}^{t_\mathrm{end}} \mathrm{d} t\, 
        H \mathrm{Re} \left[ 
            1 - \sqrt{1 - \left( \frac{2 m_{\delta \sigma}}{3H} \right)^2}
        \right]
    \right)
    \ ,
\end{align}
where $t_\mathrm{end}$ is the cosmic time at the end of inflation.%
\footnote{%
    Since we are interested in the power spectrum of $\delta \theta$, we neglect the phase factor of $\exp(-\frac{3}{2}i\int{\rm d}t\,H\sqrt{(2m_{\delta\sigma}/(3H))^2-1})$ arising for $m_{\delta\sigma}\gg H$.
} 
The effective mass of $\delta \sigma$ is given by $m_{\delta \sigma}^2 \gtrsim \lambda v^2/4$ for $t \lesssim t_\mathrm{peak}$ and $m_{\delta \sigma}^2 \simeq \sqrt{2} \epsilon v^2$ for $t \gtrsim t_\mathrm{peak}$.
While we assumed $\lambda v^2 \gg H^2$, we set $\epsilon \ll 1$ so that $\epsilon v^2 \ll H^2$.
Then, the fluctuations of $\delta \theta$ is suppressed only for $t_k < t < t_\mathrm{peak}$.
Consequently, the suppression for $k < k_\mathrm{peak}$ is approximated by
\begin{align}
    \frac{\delta \theta(t_\mathrm{end})}{\delta \theta(t_k)}
    \simeq
    \exp \left[ 
        - \frac{3}{2} (N_\mathrm{peak} - N_k)
    \right]
    \simeq
    \left( 
        \frac{k}{k_\mathrm{peak}}
    \right)^{3/2}
    \ ,
\end{align}
where $k_\mathrm{peak}$ is the wavenumber that exits the horizon at $t = t_\mathrm{peak}$, and $N_\mathrm{peak}$ and $N_k$ are the e-folding numbers corresponding to $t = t_\mathrm{peak}$ and $t_k$, respectively.

As a result, the power spectrum of $\delta \sigma$ at the end of inflation is given by 
\begin{align}
    \mathcal{P}_{\delta \sigma}(k, t_\mathrm{end})
    =
    \left( \frac{H}{2\pi} \right)^2
    \left( \frac{\delta \theta(t_\mathrm{end})}{\delta \theta(t_k)} \right)^2
    \left( \frac{v}{\varphi_0(t_k)} \right)^2
    \simeq 
    \begin{cases}
        \left( \dfrac{H}{2\pi} \right)^2
        \left( 
            \dfrac{k}{k_\mathrm{peak}}
        \right)^3
        \left( 
            \dfrac{\lambda}{4 \sqrt{2} \epsilon}
        \right)^2
        &
        (k \lesssim k_\mathrm{peak})
        \vspace{3mm} \\
        \left( \dfrac{H}{2\pi} \right)^2
        \left( 
            \dfrac{v}{\varphi_0(t_k)}
        \right)^2
        &
        (k \gtrsim k_\mathrm{peak})
    \end{cases}
    \ ,
\end{align}
where we used $\varphi_0(t_\mathrm{end}) \simeq v$.
Since $\varphi_0$ approximately grows as $v^2 - 4gI^2/\lambda$ for $t > t_\mathrm{peak}$, $\mathcal{P}_{\delta \sigma}$ is blue-tilted for $k \lesssim k_\mathrm{peak}$ and red-tilted for $k \gtrsim k_\mathrm{peak}$, and thus has a peak at $k \simeq k_\mathrm{peak}$.
On the other hand, $\mathcal{P}_{\delta \varphi}$ is highly suppressed due to its effective mass $m_{\delta \varphi} = \mathcal{O}(\lambda v^2) \gg H^2$.
Consequently, $\Phi$ effectively fluctuates only in the phase direction, $\delta \sigma$.

After inflation, $\delta \sigma$ plays the role of a curvaton field.
We assume that $\delta \sigma$ obtains an axionlike potential via some non-perturbative effects
\begin{align}
    V_\sigma
    =
    \Lambda^4 \left[ 
        1 - \cos \left( \frac{\delta \sigma}{v} + \theta_i \right)
    \right] 
    \ ,
    \label{eq: V sigma}
\end{align}
where $\Lambda$ is the scale of the potential, and $\theta_i$ is a misalignment angle between $V_\sigma$ and the $\epsilon$ term in $V_\Phi$.
We redefine the curvaton field by $\sigma \equiv v \theta_i + \delta \sigma$ and $\theta \equiv \theta_i + \delta \theta$.
Then, the potential can be rewritten as
\begin{align}
    V_\sigma
    =
    \Lambda^4 \left[ 
        1 - \cos \left( \frac{\sigma}{v} \right)
    \right] 
    \simeq 
    \frac{1}{2} m_\sigma^2 \sigma^2
    \ ,
\end{align}
where $m_\sigma \equiv \Lambda^2/v$ and we assumed $\sigma \lesssim v$.
Then, $\delta \sigma$ at the end of inflation results in the density perturbation of the curvaton field as
\begin{align}
    \frac{\delta \rho_\sigma}{\rho_\sigma}
    =
    2 \frac{\delta \theta}{\theta_i}
    \ .
\end{align}

Let us evaluate the curvature perturbations $\zeta$.
The curvature perturbations are given by 
\begin{align}
    \zeta 
    =
    - \frac{H}{\dot{\rho}} \delta \rho
    =
    - \frac{H}{\dot{\rho}_I + \dot{\rho}_\sigma}
    \left( \delta \rho_I + \delta \rho_\sigma \right)
    \ ,
\end{align}
where $\rho_I$ and $\rho_\sigma$ are the energy density of radiation from inflaton decay and the curvaton field, respectively, and $\delta \rho_I$ and $\delta \rho_\sigma$ are their fluctuations.
The dots denote derivatives with respect to the cosmic time, $t$.
Here, we consider that the decay product from the inflaton behaves as radiation while the curvaton field oscillates in $V_\sigma$ and behaves as non-relativistic matter.
Then, we obtain $\dot{\rho}_I = -4H \rho_I$, and $\dot{\rho}_\sigma = -3H \rho_\sigma$, leading to
\begin{align}
    \label{eq:zeta}
    \zeta 
    = 
    \frac{1}{4 + 3r}
    \left[
        \frac{\delta \rho_I}{\rho_I} 
        + r \frac{\delta \rho_\sigma}{\rho_\sigma}
    \right]
    \ ,
\end{align}
where $r$ is the energy ratio of the curvaton and inflaton, $\rho_\sigma/\rho_I$.%
\footnote{%
    In some literature, e.g, Ref.~\cite{Lyth:2001nq}, $r$ is defined by $r\equiv 3\rho_\sigma/(4 \rho_I + 3 \rho_\sigma)$ rather than our $r \equiv \rho_\sigma/\rho_I$.
}
Notice that $\delta\rho_\sigma$ contains the contribution from the curvature perturbation induced by the inflaton.
Thus, $\delta\rho_\sigma/\rho_\sigma$ is written as
\begin{equation}
    \frac{\delta\rho_\sigma}{\rho_\sigma}
    = \left.\frac{\delta\rho_\sigma}{\rho_\sigma}\right|_\mathrm{curv}
    +\left.\frac{\delta\rho_\sigma}{\rho_\sigma}\right|_\mathrm{inf},
\end{equation}
where the first (second) term is proportional to fluctuations of the inflaton (curvaton) during inflation.
Using the adiabatic condition ($\delta\rho_\sigma/(3\rho_\sigma)|_\mathrm{inf}=\delta\rho_I/(4\rho_I)$), the contribution from the inflaton fluctuations is given by
\begin{equation}
    \left.\frac{\delta\rho_\sigma}{\rho_\sigma}\right|_\mathrm{inf}
    = \frac{3}{4} \frac{\delta\rho_I}{\rho_I}.
\end{equation}
Thus, Eq.~\eqref{eq:zeta} is written as
\begin{align}
    \zeta = \frac{1}{4}\frac{\delta\rho_I}{\rho_{I}}
    + \frac{r}{4+3r}\left.\frac{\delta\rho_\sigma}{\rho_\sigma}\right|_\mathrm{curv}.
\end{align}
The first term is the same as the curvature perturbation when the curvaton is absent.
Hereafter, for simplicity, we denote $\delta\rho_\sigma/\rho_\sigma|_\mathrm{curv} = \delta\rho_\sigma/\rho_\sigma$.
For $k < k_\mathrm{dec}$ with $k_\mathrm{dec}$ being the scale reentering the horizon at the decay of the curvaton, the power spectrum of $\zeta$ is given by
\begin{align}
    \mathcal{P}_\zeta(k,t_\mathrm{dec})
    =
    \left( \frac{2r_\mathrm{dec}}{4+3r_\mathrm{dec}} \right)^2
    \frac{\mathcal{P}_{\delta\sigma}(k,t_\mathrm{end})}{v^2 \theta_i^2}
    + \mathcal{P}_{\zeta,I}
    \ ,
    \label{eq: P zeta of P sigma}
\end{align}
where $r_\mathrm{dec}$ is $r$ at $t = t_\mathrm{dec}$, and $\mathcal{P}_{\zeta,I}$ is the contribution from the fluctuations of the inflaton.
Since $\mathcal{P}_{\delta \sigma}$ is highly suppressed on large scales, $\mathcal{P}_{\zeta,I}$ is responsible for the curvature perturbations observed in the CMB anisotropies.
In the following, we consider the case where the mode with $k_\mathrm{peak}$ reenters the horizon well after the curvaton decay.
Then, we neglect the modes that reenter the horizon before the curvaton decay since their contribution to the PBH formation is negligibly small.
Note that such modes cannot be neglected in the type I model as discussed later.

\begin{figure}[t]
    \centering
    \includegraphics[width=.48\textwidth]{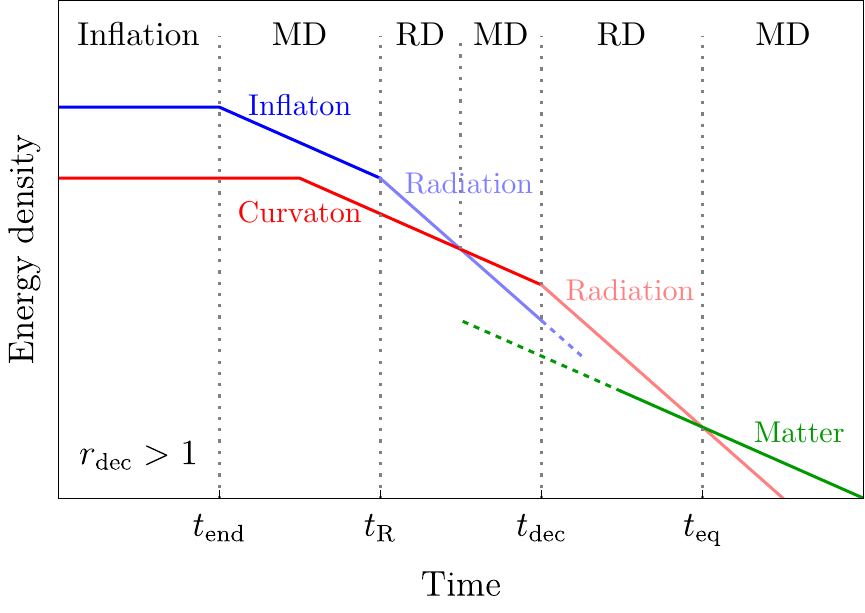}
    \hspace{5mm}
    \includegraphics[width=.48\textwidth]{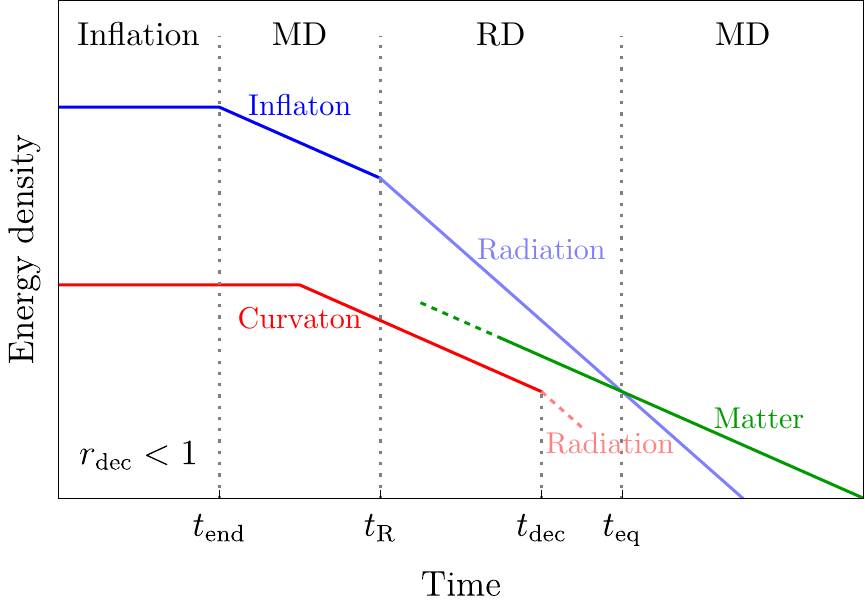}
    \caption{%
        Left:
        Evolution of the energy densities for $r_\mathrm{dec} > 1$.
        The energy density of the inflaton (blue) remains almost constant during inflation, and becomes matter-like after the end of inflation at $t = t_\mathrm{end}$.
        Then, the inflaton decays to radiation (light blue) at $t_\mathrm{R}$.
        The curvaton (red) also remains almost constant in the early stage and starts to oscillate during the MD era dominated by the oscillating inflaton.
        Then, the oscillating curvaton comes to dominate the universe and decays into radiation (light red) at $t = t_\mathrm{dec}$.
        Finally, non-relativistic matter (green) dominates the universe for $t > t_\mathrm{eq}$ as in the standard cosmology.
        Right: 
        The same for $r_\mathrm{dec} < 1$.
        In this case, the curvaton never dominates the universe, and there is a single RD era.
    }
    \label{fig: Energy}
\end{figure}

Finally, to evaluate the primordial power spectrum, we need to relate the scales during inflation to those after inflation.
The former includes $k_\mathrm{peak}$, and the latter includes the horizon scale at the matter-radiation equality in the standard cosmology, $k_\mathrm{eq} = 0.0103\,\mathrm{Mpc}^{-1}$.
In this scenario, the inflation ends at $t = t_\mathrm{end}$, and the early matter-dominated (MD) era follows due to the oscillating energy of the inflaton.
We assume that the curvaton starts to oscillate during this early MD era.
The inflaton decays at $t = t_\mathrm{R}$, leading to the first radiation-dominated (RD) era.
Then, if $r_\mathrm{dec} > 1$, the oscillation of the curvaton dominates the universe, leading to the MD era.
Finally, the curvaton decays at $t = t_\mathrm{dec}$, and the universe is dominated by radiation as in the standard cosmology.
After that, the universe experiences the matter-radiation equality at $t = t_\mathrm{eq}$.
For simplicity, we assume that the inflaton and curvaton instantaneously decay at $t = t_\mathrm{R}$ and $t_\mathrm{dec}$, respectively.
If $r_\mathrm{dec} < 1$, the curvaton does not dominate the universe, and there is a single RD era after inflation.
We show the time evolution of the energy densities of each component in Fig.~\ref{fig: Energy}.

Consequently, we obtain the relation as~\cite{Inomata:2023drn}
\begin{align}
    \frac{k}{k_\mathrm{eq}}
    \equiv 
    1.3 \times 10^{24} 
    \left( 1 + r_\mathrm{dec} \right)^{-1/4}
    \left( \frac{\rho^\mathrm{tot}_\mathrm{end}}{3 \times 10^{-11} M_\mathrm{Pl}^4} \right)^{1/6}
    \left( \frac{T_\mathrm{R}}{10^{14}\,\mathrm{GeV}} \right)^{1/3}
    \left( \frac{g_\mathrm{R}}{g^s_\mathrm{R}} \right)^{1/3}
    \left( \frac{g_\mathrm{dec,aft}}{g_\mathrm{dec,bef}} \right)^{1/4}
    \left( \frac{g^{s}_\mathrm{dec,aft}}
    {g^{s}_\mathrm{dec,bef}} \right)^{-1/3}
    \frac{k}{\mathcal{H}_\mathrm{end}}
    \ ,
    \label{eq: k relation}
\end{align}
where $\rho^\mathrm{tot}$ is the total energy density of the universe, $g$ and $g^s$ are the effective degrees of freedom of relativistic species for the energy and entropy densities, respectively, $\mathcal{H} \equiv a H$ is the conformal Hubble parameter, and the quantities with subscripts correspond to those at the time with the same subscripts.
The additional subscripts, bef and aft, denote the quantities just before and after $t = t_\mathrm{dec}$, respectively.
We present the derivation of this relation in Appendix~\ref{app: wavenumber}.
Note that this relation applies for arbitrary $r_\mathrm{dec}$.

To be concrete, we adopt the Starobinsky-type potential for the inflaton potential~\cite{Starobinsky:1980te}:
\begin{align}
    V_I(I)
    =
    \frac{3}{4} M_\mathrm{Pl}^2 M^2 
    \left[
        1 - \exp \left(
            -\sqrt{\frac{2}{3}} \frac{I}{M_\mathrm{Pl}}
        \right)
    \right]^2
    \ ,
\end{align}
where $M_\mathrm{Pl} \simeq 2.4 \times 10^{18}$\,GeV is the reduced Planck mass, and we take $M = 1.23 \times 10^{-5} M_\mathrm{Pl}$, which is consistent with the Planck results~\cite{Planck:2018vyg,Planck:2018jri}.
Note that the corresponding tensor-to-scalar ratio on the CMB scales is given by $r_t \sim 4 \times 10^{-3}$, which is well below the current upper limit~\cite{Galloni:2024lre}.
We consider three sets of the parameters as summarized in Table~\ref{tab: type II parameter}.
\begin{table*}[htbp]
    \caption{Model parameters for the type II axionlike curvaton model. 
    The colors denote the line colors used in Figs.~\ref{fig: Pzeta}, \ref{fig: fPBH}, and \ref{fig: OmegaGW}.}
    \label{tab: type II parameter}
    \begin{tabular}{ c | c | c | c | c | c | c | c | c}
         $r_\mathrm{dec}$ & $v/M_\mathrm{Pl}$ & $\epsilon$ & $g$ & $\theta_i$ & $\lambda$ & $T_\mathrm{R}$\,[GeV] & $M/M_\mathrm{Pl}$ & color
        \\
        \hline
        $3$ & $0.025$ & $9.5 \times 10^{-10}$ & $6.35 \times 10^{-10}$ & $0.00199$ & $8.4 \times 10^{-5}$ & $10^{14}$ & $1.23 \times 10^{-5}$ & blue
        \\
        $0.3$ & $0.025$ & $1.7 \times 10^{-9}$ & $6.13 \times 10^{-10}$ & $0.00122$ & $8.4 \times 10^{-5}$ & $10^{14}$ & $1.23 \times 10^{-5}$ & red
        \\
        $0.3$ & $0.0235$ & $9.0 \times 10^{-10}$ & $6.10 \times 10^{-10}$ & $0.00536$ & $8.4 \times 10^{-5}$ & $10^{14}$ & $1.23 \times 10^{-5}$ & violet
    \end{tabular}
\end{table*}

\begin{figure}[t]
    \centering
    \includegraphics[width=.8\textwidth]{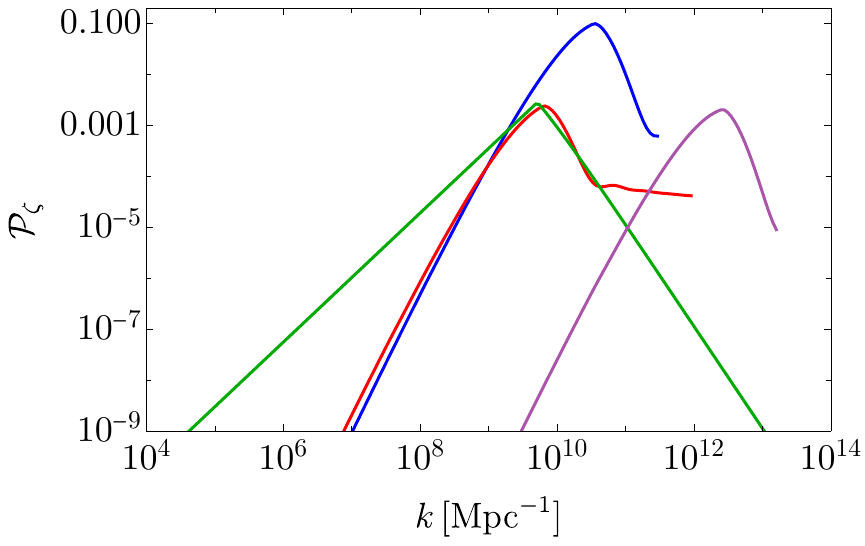}
    \caption{%
        Power spectra of the curvature perturbations, $\mathcal{P}_\zeta$.
        The blue, red, and violet lines represent the results for the type II model with the parameter sets in Table~\ref{tab: type II parameter}.
        The curvature perturbations have peaks due to the evolution of $\varphi_0$ during inflation.
        Due to the effective mass of $\delta \sigma$ in the early stage of inflation, they are suppressed for small $k$. 
        We do not show the short modes that reenter the horizon before the curvaton decay.
        While the red and blue lines lead to the production of PBHs explaining the microlensing events, the violet line corresponds to PBHs with smaller masses.
        The green line represents the result for the type I model.
    }
    \label{fig: Pzeta}
\end{figure}

We show the power spectra of the curvature perturbations in Fig.~\ref{fig: Pzeta}.
The blue and red lines represent the results for the parameter sets with $(r_\mathrm{dec},~v/M_\mathrm{Pl}) = (3,~0.025)$ and  $(0.3, 0.025)$, respectively.
As shown later, these power spectra lead to the production of PBHs that account for the microlensing events.
We also show the power spectrum  (violet line) for the parameter set with $(r_\mathrm{dec},~v/M_\mathrm{Pl}) = (3,~0.0235)$, which produces PBHs with smaller masses $\sim 10^{-12}\,M_\odot$ (see Fig.~\ref{fig: fPBH}).
As mentioned above, we focus on the modes that reenter the horizon after the curvaton decay, and the power spectra for larger $k$ are not shown.

\section{Type I axionlike curvaton model}
\label{sec: type I}

Next, we review the type I axionlike curvaton model~\cite{Kawasaki:2012wr,Kawasaki:2013xsa,Bugaev:2013vba,Ando:2017veq,Inomata:2020xad,Ferrante:2023bgz}.
We mainly follow Refs.~\cite{Ando:2017veq,Inomata:2020xad} while the effects of non-instantaneous decay of the curvaton and higher-order non-Gaussianity of the curvature perturbations are discussed in Ref.~\cite{Ferrante:2023bgz}.
In this scenario, we consider that a complex scalar field $\Phi$ is decoupled from the inflaton $I$, and the Lagrangian is given by 
\begin{align}
    \mathcal{L}^\mathrm{(I)}
    =
    \frac{1}{2} \partial_\mu I \partial^\mu I 
    + \partial_\mu \Phi \partial^\mu \Phi^*
    - V_I(I) - V^\mathrm{(I)}_\Phi(\Phi)
    \ ,
\end{align}
where $V_\Phi^{\mathrm{(I)}}$ is the potential for $\Phi$.
The type I setup is realized with another scalar field in the framework of supersymmetry~\cite{Kasuya:1996ns,Kasuya:2009up} (see Appendix~\ref{app: type_I_model}).
While $|\Phi|$ takes a nonzero value at the potential minimum, $ V^\mathrm{(I)}_\Phi$ can be approximated for $|\Phi| \gg v$ by 
\begin{align}
    V^\mathrm{(I)}_\Phi(\Phi)
    =
    c H^2 |\Phi|^2
    \ ,
\end{align}
where $c$ is a dimensionless coupling constant.
We define the VEV of $\Phi$ by $|\Phi| = v/\sqrt{2}$ again.
In addition, we assume $V_\sigma$ given by Eq.~\eqref{eq: V sigma} as in the type II model.

In this scenario, we assume the initial condition of $\varphi_0 \gg v$.
Using the decomposition of Eq.~\eqref{eq: Phi decomposition}, we obtain the equation of motion for the radial component as
\begin{align}
    \ddot{\varphi}_0 + 3 H \dot{\varphi}_0 
    + c H^2 (\varphi_0 - v)
    =
    0
    \ ,
\end{align}
where we adopt the effective potential of $c H^2 (\varphi_0-v)^2/2$ for $\varphi_0$ taking into account that $V_\Phi^\mathrm{(I)}$ is minimized at $\varphi = 0$ (see Appendix~\ref{app: type_I_model}).
Assuming $H = \mathrm{const}.$ and $\varphi_0 \gg v$, we obtain the approximate solution of 
\begin{align}
    \varphi_0(t) 
    \simeq 
    \begin{cases}
        v e^{- \lambda H (t - t_*)}
        &
        (t < t_*)
        \vspace{1mm} \\
        v
        &
        (t \geq t_*)
    \end{cases}
    \ ,
\end{align}
where $t_*$ is the time when $\varphi$ settles to the potential minimum, and $\lambda$ is given by 
\begin{align}
    \lambda 
    =
    \frac{3}{2} \left( 1 - \sqrt{1 - \frac{4c}{9}} \right)
    \ .
\end{align}

As a result, the power spectrum of $\delta \sigma$ at the end of inflation is given by 
\begin{align}
    \mathcal{P}_{\delta \sigma}(k, t_\mathrm{end})
    =
    \left( \frac{H}{2\pi} \right)^2
    \left( \frac{v}{\varphi_0(t_k)} \right)^2
    \simeq 
    \begin{cases}
        \left( \dfrac{H}{2\pi} \right)^2
        \left( 
            \dfrac{v}{\varphi_0(t_k)}
        \right)^2
        &
        (k \lesssim k_*)
        \vspace{3mm} \\
        \left( \dfrac{H}{2\pi} \right)^2
        &
        (k \gtrsim k_*)
    \end{cases}
    \ ,
\end{align}
where $k_*$ is the wavenumber that exits the horizon at $t = t_*$.
From Eq.~\eqref{eq: P zeta of P sigma}, we obtain the power spectrum of the curvature perturbation.
If the contribution from the curvaton is dominant, $\mathcal{P}_\zeta$ is proportional to $k^{2\lambda}$ for $k \lesssim k_*$ and scale invariant for $k \gtrsim k_*$.
Since $\lambda > 0$, $\mathcal{P}_\zeta$ becomes blue-tilted for $k \lesssim k_*$.

Since PBHs are formed when the density fluctuations reenter the horizon, we should use the curvature perturbation at the horizon reentry to evaluate the PBH abundance.
We show the power spectrum of the curvature perturbations at the horizon reentry of each $k$ with the green line in Fig.~\ref{fig: Pzeta}.
Here, we adopt the following parameters:
\begin{equation}
\begin{gathered}
    r_\mathrm{dec} = 0.3 
    \ , \quad 
    v/M_\mathrm{Pl} = 0.001
    \ , \quad 
    c = 1.5
    \ , \quad 
    k_* = 5 \times 10^9\,\mathrm{Mpc}^{-1}
    \ ,
    \\
    \theta_i = 0.0023
    \ , \quad 
    T_\mathrm{R} = 10^{14}\,\mathrm{GeV}
    \ , \quad 
    M/M_\mathrm{Pl} = 1.26 \times 10^{-5}
    \ .
\end{gathered}
\label{eq: type I parameters}
\end{equation}
After the inflaton decay and before the curvaton decay, $r$ evolves as $r(t) = r_\mathrm{dec} a(t)/a_\mathrm{dec}$ for constant $g$ and $g^s$.
Thus, we evaluate $\mathcal{P}_\zeta$ using $r(t_k)$ instead of $r_\mathrm{dec}$ in Eq.~\eqref{eq: P zeta of P sigma} for $k > k_\mathrm{dec} \simeq 5.3 \times10^9\,\mathrm{Mpc}^{-1}$.
On the other hand, the production of gravitational waves is governed by the gravitational potential, $\Psi$.
To evaluate the gravitational wave abundance in the type I model, we need to follow the evolution of $\Psi$ taking into account isocurvature perturbations before the curvaton decay, which we will discuss in Sec.~\ref{sec: GW}.

\section{PBH formation from non-Gaussian curvature perturbations}
\label{sec: PBH}

As we have seen above, the axionlike curvaton models predict the enhancement of the curvature perturbations on small scales.
In addition, the curvaton scenarios induce local-type non-Gaussianity in curvature perturbations, which affects the PBH formation rate.
In the following, we evaluate the PBH formation in our scenario following Refs.~\cite{Kawasaki:2019mbl,Inomata:2023drn}.

In the curvaton scenario, the non-Gaussianity of the curvature perturbations can be represented by the local-type non-Gaussianity as
\begin{align}
    \zeta(x)
    =
    \zeta_g(x) + \frac{3}{5} f_\mathrm{NL} \left( \zeta^2_g(x) - \langle \zeta^2_g(x) \rangle \right)
    \ ,
\end{align}
where $\zeta_g$ follows the Gaussian statistics.
The non-Gaussianity parameter $f_\mathrm{NL}$ is related to $r_\mathrm{dec}$ as~\cite{Ando:2017veq}
\begin{align}
    f_\mathrm{NL}
    =
    \frac{5}{12} \left( 
        -3 + \frac{4}{r_\mathrm{dec}} + \frac{8}{4 + 3r_\mathrm{dec}}
    \right)
    \ .
\end{align}

PBHs are formed when the coarse-grained density fluctuation exceeds a threshold value at the horizon reentry.
Due to the non-linear relation between the density and curvature perturbations, the density perturbation is represented using the Gaussian density perturbations, $\delta_g$ as~\cite{Kawasaki:2019mbl}
\begin{align}
    \delta(R)
    =
    \delta_g(R)
    + \frac{\mu_3(R)}{6\sigma(R)} 
    \left( \delta_g^2(R) - \sigma^2(R) \right)
    \ ,
\end{align}
where $R$ is the smoothing scale, and $\sigma^2$ and $\mu_3$ are the variance and skewness of the density perturbations, respectively.
The smoothing length is related to the wavenumber of the PBH scale as $k_\mathrm{PBH} = 1/R$. 
Neglecting higher-order terms, $\sigma^2$ coincides with the variance of $\delta_g$. 
In other words, $\delta_g(R)$ follows the Gaussian probability distribution of
\begin{align}
    P_g(\delta_g;R)
    =
    \frac{1}{\sqrt{2\pi}\sigma(R)}
    \exp \left[ -\frac{\delta_g^2}{2 \sigma^2(R)} \right]
    \ .
\end{align}
The variance $\sigma^2(R)$ is obtained from the relation at leading order:
\begin{align}
    \delta_k
    =
    \frac{4}{9} \left( \frac{k}{aH} \right)^2 \zeta_k
    \ .
\end{align}
The coarse-grained density fluctuation is defined by
\begin{align}
    \delta(R)
    \equiv 
    \int \frac{\mathrm{d}^3 k}{(2\pi)^3} \,
    e^{i \bm{k}\cdot \bm{x}} \tilde{W}(k; R) \delta_k 
    \ ,
\end{align}
where $\tilde{W}(k; R)$ is the window functions in the momentum space.
Then, $\sigma^2(R)$ at $\eta$ is given by 
\begin{align}
    \sigma^2(R)
    \equiv 
    \langle \delta(R)^2  \rangle 
    &=
    \int \frac{\mathrm{d}^3 q_1}{(2\pi)^3}
    \frac{\mathrm{d}^3 q_2}{(2\pi)^3} \,
    e^{i \bm{x} \cdot(\bm{q}_1 + \bm{q}_2)} 
    \tilde{W}(q_1; R) \tilde{W}(q_2; R)
    \langle \delta_{q_1}(\eta) \delta_{q_2}(\eta) \rangle
    \nonumber\\ 
    &=
    \int \frac{\mathrm{d}^3 q}{(2\pi)^3} \,
    \tilde{W}(q; R)^2
    \frac{2\pi^2}{q^3} \mathcal{P}_\delta(q) 
    T(q\eta)^2
    \nonumber\\ 
    &=
    \frac{16}{81} 
    \int \mathrm{d} \ln q \,
    \tilde{W}(q; R)^2
    \left( q R \right)^4
    \mathcal{P}_\zeta(q) 
    T(q\eta)^2
    \ ,
\end{align}
where $T(q,\eta)$ is the transfer function.
In the radiation-dominated era, $T(q\eta)$ is given by 
\begin{align}
    T(q\eta)
    =
    3 \frac{\sin\left(\frac{q\eta}{\sqrt{3}}\right) - \frac{q\eta}{\sqrt{3}} \cos\left(\frac{q\eta}{\sqrt{3}}\right)}
    {\left(\frac{q\eta}{\sqrt{3}}\right)^3}
    \ .
\end{align}
As for the window function, we adopt the real-space top hat function, which corresponds to evaluating the spatially averaged density fluctuation at a certain radius:
\begin{align}
    \tilde{W}(q; R)
    =
    3 \frac{\sin (q R) - q R \cos(q R)}{(q R)^3}
    \ .
\end{align}
The skewness, $\mu_3(R)$ is given by
\begin{align}
    \frac{\mu_3(R)}{\sigma(R)}
    =
    Q(R) f_\mathrm{NL}
    - \frac{9}{4}
    \ ,
\end{align}
Here, the first term comes from the non-Gaussianity of the curvature perturbations, and the second term comes from the non-linear relation between the density and curvature perturbations.
The coefficient $Q(R)$ is given by 
\begin{align}
    Q(R)
    =&
    \frac{128}{405 \sigma^4(R)}
    \int \frac{\mathrm{d}p}{p}\tilde{W}(p R) 
    (p R)^2 \mathcal{P}_\zeta(p) T(p R)^2
    \int \frac{\mathrm{d}q}{q}\tilde{W}(q R) 
    (q R)^2 \mathcal{P}_\zeta(q) T(q R)^2
    \nonumber \\
    & \times 
    \int_{-1}^1 \frac{\mathrm{d}\cos \theta}{2}
    \tilde{W}(|\bm{p} + \bm{q}| R) 
    (|\bm{p} + \bm{q}| R)^2
    \ .
\end{align}
After some calculations, we obtain
\begin{align}
    \sigma^2(R)
    &=
    \frac{16}{27} \left[ 
        \mathcal{I}_s(R)
        -
        \mathcal{I}_c(R)
    \right]
    \ , \\
    Q(R)
    &=
    \frac{256}{135}
    \frac{\mathcal{I}_s(R)}{\sigma^4(R)} 
    \left[ 
        \mathcal{I}_s(R)
        -
        \mathcal{I}_c(R)
    \right]
    \ ,
\end{align}
with 
\begin{align}
    \mathcal{I}_s(R)
    &\equiv 
    \int \frac{\mathrm{d}x}{x}\tilde{W}(x) 
    \mathcal{P}_\zeta(x/R) T(x)^2 x \sin x
    \ , \\
    \mathcal{I}_c(R)
    &\equiv 
    \int \frac{\mathrm{d}x}{x}\tilde{W}(x) 
    \mathcal{P}_\zeta(x/R) T(x)^2 x^2 \cos x
    \ ,
\end{align}
In the Press-Schechter formalism~\cite{Press:1973iz},%
\footnote{%
    The PBH formation rate is also discussed using the peak theory~\cite{Bardeen:1985tr} even for the non-Gaussian case~\cite{Yoo:2019pma,Kitajima:2021fpq,Young:2022phe,Inui:2024fgk}.
    In the peak theory, the suppression and enhancement of the PBH formation are more significant than in the Press-Schechter formalism~\cite{Kitajima:2021fpq}.
    However, $\mathcal{P}_\zeta$ to produce a fixed PBH abundance is of a similar order in both procedures.
    Thus, our scenario can produce the candidate PBHs for the microlensing events since $\mathcal{P}_\zeta$ can be easily modified by changing the choice of the model parameters.
}
the PBH production rate is given by
\begin{align}
    \beta(R)
    &=
    \int_\mathrm{-\infty}^\infty 
    \mathrm{d} \delta_g \,
    P_g(\delta_g;R) \Theta(\delta(\delta_g;R) - \delta_\mathrm{th})
    \nonumber \\
    &=
    \begin{dcases}
        \int_{\delta_{g,\mathrm{th}}^{(+)}}^\infty 
        \mathrm{d} \delta_g \, P_g(\delta_g;R) 
        +
        \int_{-\infty}^{\delta_{g,\mathrm{th}}^{(-)}}
        \mathrm{d} \delta_g \, P_g(\delta_g;R) 
        &
        (\mu_3 > 0)
        \\
        \int_{\delta_{g,\mathrm{th}}^{(+)}}^{\delta_{g,\mathrm{th}}^{(-)}} 
        \mathrm{d} \delta_g \, P_g(\delta_g;R) 
        &
        (\mu_3 < 0)
    \end{dcases}
    \ ,
\end{align}
where $\delta_\mathrm{th}$ is the threshold value for PBH formation, and $\delta_{g,\mathrm{th}}^{(\pm)}$ are given by 
\begin{align}
    \delta_{g,\mathrm{th}}^{(\pm)}
    =
    \frac{3 \sigma}{\mu_3}
    \left( 
        - 1 \pm
        \sqrt{1 + \frac{2\mu_3}{3} \left( \frac{\mu_3}{6} + \frac{\delta_\mathrm{th}}{\sigma} \right) }~
    \right)
    \ .
\end{align}
In the following, we adopt $\delta_\mathrm{th} = 0.53$~\cite{Harada:2015yda} as a fiducial value in the RD era.

Before evaluating the PBH abundance, we relate the coarse-graining scale $k$ and the PBH mass (see also Appendix~\ref{app: PBH mass}).
The PBH mass is estimated by the horizon mass when the corresponding scale $k_\mathrm{PBH}$ reenters the horizon, and the PBH is formed at $t = t_\mathrm{form}$. 
Thus, we obtain~\cite{Inomata:2020xad}
\begin{align}
    M_\mathrm{PBH}
    &\simeq 
    M_\odot \frac{\gamma}{0.2} 
    \left( \frac{g_\mathrm{form}}{10.75} \right)^{1/2}
    \left( \frac{g^s_\mathrm{form}}{10.75} \right)^{-2/3}
    \left( \frac{k_\mathrm{PBH}}{2.0 \times 10^{6}\,\mathrm{Mpc}^{-1}} \right)^{-2}
    \ ,
\end{align}
where $\gamma$ is the ratio of the PBH mass to the horizon mass.
In this paper, we adopt $\gamma = 0.2$~\cite{Carr:1975qj}.

Finally, we obtain the PBH mass function as~\cite{Kawasaki:2021ycf}
\begin{align}
    f_\mathrm{PBH}(M_\mathrm{PBH}) 
    &\equiv 
    \frac{1}{\Omega_\mathrm{DM}}
    \frac{\mathrm{d} \Omega_\mathrm{PBH}}{\mathrm{d} \ln M_\mathrm{PBH}}
    \nonumber \\
    &= 
    \frac{\beta(M_\mathrm{PBH})}{1.7 \times 10^{-8}}
    \left( \frac{\gamma}{0.2} \right)^{3/2}
    \left( \frac{g_\mathrm{form}}{10.75} \right)^{3/4}
    \left( \frac{g^s_\mathrm{form}}{10.75} \right)^{-1}
    \left( \frac{\Omega_\mathrm{c}h^2}{0.12} \right)^{-1}
    \left( \frac{M_\mathrm{PBH}}{M_\odot} \right)^{-1/2}
    \ ,
\end{align}
where $\Omega_\mathrm{PBH}$ and $\Omega_\mathrm{c}$ are the density parameters of PBHs and dark matter, respectively.
See Appendix~\ref{app: PBH mass} for its derivation.

\begin{figure}[t]
    \centering
    \includegraphics[width=.8\textwidth]{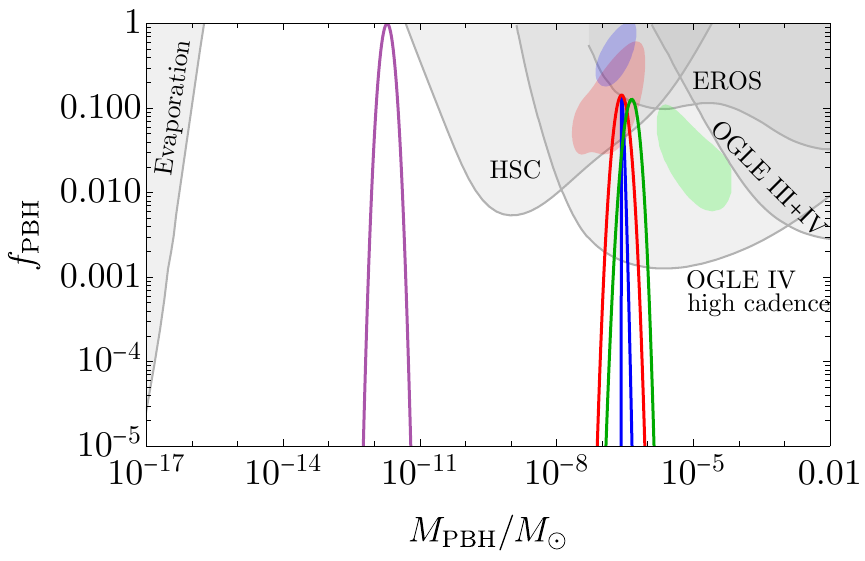}
    \caption{%
        PBH fraction to dark matter, $f_\mathrm{PBH}$, for the type II model with the parameter sets in Table~\ref{tab: type II parameter} (blue, red, and violet) and the type I model with the parameter set in Eq.~\eqref{eq: type I parameters} (green).
        The blue- and red-shaded regions represent the PBH interpretation of the 12 candidates and 4 secure candidates for microlensing events observed in Subaru HSC observations~\cite{Sugiyama:2026kpv}, respectively.
        The green-shaded region represents the estimate by Ref.~\cite{Niikura:2019kqi} interpreting the ultrashort-timescale microlensing events in the 5-year OGLE data~\cite{Mroz:2017mvf} as originating from PBHs.
        The gray-shaded regions represent the observational upper bounds on the PBH abundance by EROS~\cite{EROS-2:2006ryy}, HSC~\cite{Croon:2020ouk}, OGLE-III+OGLE-IV~\cite{Mroz:2024mse}, the high-cadence survey by OGLE-IV~\cite{Mroz:2024wia}, X-ray background~\cite{Tan:2024nbx}, and the 21\,cm measurement~\cite{Mittal:2021egv}.
        The last two limits are due to the evaporation of PBHs and are taken from PBHbounds~\cite{PBHbounds}.
    }
    \label{fig: fPBH}
\end{figure}

In Fig.~\ref{fig: fPBH}, we show the PBH mass functions for the power spectra of the curvature perturbations shown in Fig.~\ref{fig: Pzeta}.
The blue, red, violet, and green lines represent each case as in Fig.~\ref{fig: Pzeta}.
The gray regions are limited by Expérience de Recherche d'Objets Sombres (EROS)~\cite{EROS-2:2006ryy}, Hyper Suprime-Cam (HSC)~\cite{Croon:2020ouk}, Optical Gravitational Lensing Experiment(OGLE)~\cite{Mroz:2024mse}, the high-cadence survey by OGLE~\cite{Mroz:2024wia}, X-ray background~\cite{Tan:2024nbx}, and the 21\,cm measurement~\cite{Mittal:2021egv}.
It is seen that both type I and II axionlike curvaton models can produce PBHs with mass $M_\mathrm{PBH}\sim 10^{-6}\,M_\odot$ and abundance $f_\mathrm{PBH} \sim 0.1$, which explains the HSC microlensing events~\cite{Sugiyama:2026kpv}.
Moreover, the type II model can also produce PBHs with mass $\sim 10^{-12}\,M_\odot$ and account for all the dark matter of the universe.

In Fig.~\ref{fig: fPBH}, we see that the limit by the high cadence survey of OGLE~\cite{Mroz:2024wia} is in tension with the favored regions by the HSC observation.
Ref.~\cite{Sugiyama:2026kpv} discusses possible reasons for this tension including systematics in the detection efficiency, but it may need further study to resolve the tension.

For the type II models, we set $\mathcal{P}_\zeta = 0$ for $k > k_\mathrm{dec}$.
As mentioned above, the curvature perturbations on such scales are sufficiently small compared to $k \sim k_\mathrm{peak}$ and do not significantly affect the estimates of the PBH abundance.
For the type I model, the curvature perturbations for $k > k_\mathrm{dec}$ reenter the horizon before the curvaton decay, and their non-Gaussianity deviates from that evaluated with $r_\mathrm{dec}$.
However, the scales relevant to the PBH formation reenter the horizon slightly before the curvaton decay, and thus we approximate $f_\mathrm{NL}$ by that given with $r_\mathrm{dec}$ for all scales.

\section{Scalar-induced gravitational waves}
\label{sec: GW}

When the scalar perturbations are enhanced on a certain scale, gravitational waves are generated in the second order on the corresponding scale.
First, we consider the gravitational waves in the type II model, where we focus on the scales that reenter the horizon after the curvaton decay.
Then, we can use the simple relation between the curvature perturbations and gravitational potential as in the standard cosmology. 
In this case, the energy density of the gravitational waves in the subhorizon limit during the RD era is given by~\cite{Espinosa:2018eve,Kohri:2018awv}
\begin{align}
    \frac{\mathrm{d}\Omega_\mathrm{GW,RD}}{\mathrm{d}\ln k}
    =
    \int_0^\infty \mathrm{d} v \,
    \int_{|1-v|}^{1+v} \mathrm{d} u \,
    \mathcal{K}(u,v) \mathcal{P}_\zeta(u k) \mathcal{P}_\zeta(v k)
    \ ,
\end{align}
with
\begin{align}
    \mathcal{K}(u,v)
    =
    \frac{3[4v^2 - (1+v^2-u^2)^2]^2 (u^2+v^2-3)^4}{1024 u^8 v^8}
    \left[ 
        \left( 
            \ln \left| \frac{3-(u+v)^2}{3-(u-v)^2} \right|
            -
            \frac{4uv}{u^2 + v^2 - 3}
        \right)^2
        +
        \pi^2 \Theta(u + v - \sqrt{3})
    \right]
    \ .
\end{align}
Here, we neglected the effect of the non-Gaussianity of the curvature perturbations on SIGWs, which is not so large for $|f_\mathrm{NL}| \lesssim 1$~\cite{Nakama:2016gzw,Garcia-Bellido:2017aan,Cai:2018dig,Unal:2018yaa,Adshead:2021hnm,Abe:2022xur,Yuan:2023ofl,Li:2023xtl,Pi:2024lsu}.

The density parameter during the RD era is related to the current density parameter as
\begin{align}
    \frac{\mathrm{d}\Omega_{\mathrm{GW},0} h^2}{\mathrm{d}\ln k}
    =
    0.39
    \left( \frac{g_\mathrm{h}}{106.75} \right)
    \left( \frac{g^s_\mathrm{h}}{106.75} \right)^{-4/3}
    \Omega_{\mathrm{r},0} h^2
    \frac{\mathrm{d}\Omega_\mathrm{GW,RD}}{\mathrm{d}\ln k}
    \ ,
\end{align}
where $\Omega_{\mathrm{r},0}h^2 \simeq 4.2 \times 10^{-5}$~\cite{Planck:2018vyg} is the current density parameter of radiation with $h \equiv H_0/(100\,\mathrm{km/s/Mpc})$ being the reduced Hubble constant, and the subscripts, h and $0$, denote the time of the horizon reentry of the corresponding mode and the current time, respectively.

In the type I model, the modes that reenter the horizon before the curvaton decay are also enhanced as seen in Fig.~\ref{fig: Pzeta}.
Before the curvaton decay, the curvaton field induces isocurvature perturbations, which affect the evolution of the gravitational potential.
In this case, we need to solve the transfer function of the gravitational potential sourced by the fluctuations of the curvaton field.
The power spectrum of the gravitational waves is given by~\cite{Ando:2017veq} 
\begin{align}
    \mathcal{P}_h(\eta, k)
    =
    \frac{1}{4} \int_0^\infty \mathrm{d} v 
    \int_{|1 - v|}^{1+v} \mathrm{d} u \,
    \frac{v^2}{u^2}
    \left( 1 - \frac{{(1 + v^2 - u^2)}^2}{4 v^2} \right)^2
    \mathcal{P}_S (vk) \mathcal{P}_S (uk)
    \left[ 
        \frac{k^2}{a(\eta)} 
        \int^\eta \mathrm{d} \tilde{\eta} \,
        a(\tilde{\eta}) g_k(\eta; \tilde{\eta})
        f(vk, uk, \tilde{\eta})
    \right]^2
    \ ,
\end{align}
where $g_k$ is the Green function of the gravitational wave given by 
\begin{align}
    g_k(\eta; \tilde{\eta}) 
    \equiv 
    \frac{\sin [k(\eta - \tilde{\eta})]}{k}
    \ ,
\end{align}
$\mathcal{P}_S \equiv 4\mathcal{P}_{\delta \sigma}/(v^2 \theta_i^2)$ denotes the primordial power spectrum of the isocurvature perturbations, and $f$ involves the transfer function of the gravitational potential, $T(\eta)$, and is given by 
\begin{align}
    f(k_1, k_2, \eta)
    \equiv 
    4 \left[ 
        3T(k_1) T(k_2) 
        + \frac{2}{\mathcal{H}} T'(k_1) T(k_2)
        + \frac{1}{\mathcal{H}^2} T'(k_1) T'(k_2)
    \right]
    \ .
\end{align}
Here, the transfer function for $\eta < \eta_\mathrm{dec}$ can be approximated by a fitting function of~\cite{Ando:2017veq}
\begin{align}
    T(\eta, k)
    =
    - \frac{r(\eta)}{6 + 5r(\eta)}
    (1 - \Delta_\sigma(\eta))
    \left[ 
        1 + \frac{2(1+r(\eta))}{3(6+5r(\eta))}
        (1 - \Delta_\sigma(\eta))
        (k\eta)^2
    \right]^{-1}
    \ ,
\end{align}
with 
\begin{align}
    \Delta_\sigma(\eta)
    \equiv 
    1 + \frac{(6 + 5r(\eta))(-r(\eta)^3 + 2r(\eta)^2 - 8r(\eta) - 16 + 16\sqrt{1 + r(\eta)})}{5 r(\eta)^4}
    \ .
\end{align}
The current density parameter of the gravitational wave is given by 
\begin{align}
    \frac{\mathrm{d}\Omega_{\mathrm{GW},0} h^2}{\mathrm{d}\ln k}
    =
    0.83 \Omega_{\mathrm{r},0} h^2
    \left( \frac{g_\star}{10.75} \right)^{-1/3}
    \frac{k^2}{\mathcal{H}(\eta_\star)^2}
    \overline{\mathcal{P}_h(\eta_\star,k)}
    \ ,
\end{align}
where the overline represents the time average and $\eta_\star$ is a certain conformal time when the gravitational waves behave as radiation.

\begin{figure}[t]
    \centering
    \includegraphics[width=.8\textwidth]{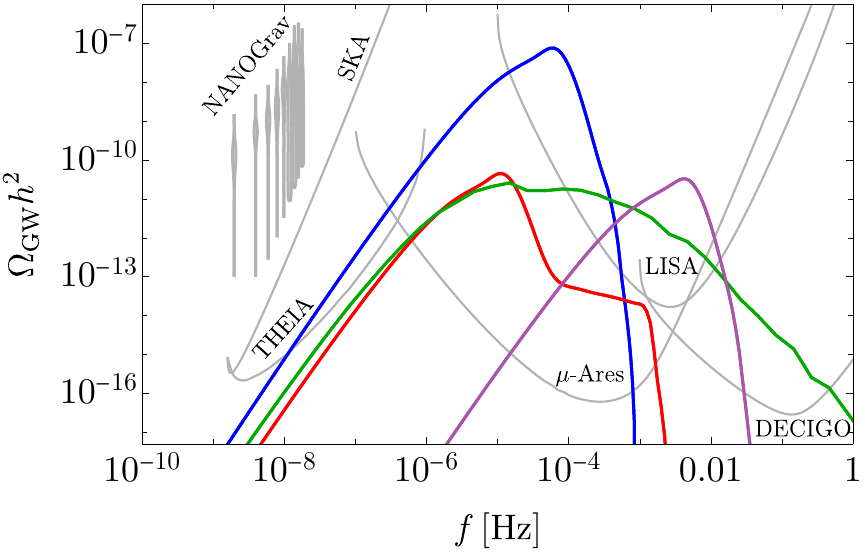}
    \caption{%
        Power spectra of the gravitational waves in the type II (blue, red, and violet) and type I (green) axionlike curvaton models.
        The colors are the same as in Figs.~\ref{fig: Pzeta} and \ref{fig: fPBH}.
        The gray lines represent the future sensitivities of the Square Kilometre Array (SKA)~\cite{Janssen:2014dka,Weltman:2018zrl}, THEIA~\cite{Garcia-Bellido:2021zgu}, $\mu$Ares~\cite{Sesana:2019vho}, Laser Interferometer Space Antenna (LISA)~\cite{LISA:2017pwj}, and DeciHertz Interferometer Gravitational-Wave Observatory (DECIGO)~\cite{Kawamura:2006up}.      
        The sensitivities of SKA, LISA, and DECIGO are taken from Ref.~\cite{Schmitz:2020syl}.
        The gray violins on the nHz scale represent the result of the NANOGrav 15\,yr data set~\cite{NANOGrav:2023gor}.
    }
    \label{fig: OmegaGW}
\end{figure}

We show the power spectrum of scalar-induced gravitational waves in Fig.~\ref{fig: OmegaGW}.
The line colors are the same as in Figs.~\ref{fig: Pzeta} and \ref{fig: fPBH}, and the frequency of the gravitational waves is related to the wavenumber by $f = k/(2\pi)$.
The power spectra have peaks at frequencies that correspond to the peaks of the curvature power spectra (see Fig.~\ref{fig: Pzeta}).
While the resulting gravitational waves are within the reach of future observations in all cases, the spectrum is largest in the type II model with $r_\mathrm{dec} = 3$, reflecting the amplitude of the curvature power spectrum.
Due to the inclusion of $k > k_\mathrm{dec}$, the high-frequency tails are larger in the type I model.

\section{Summary}
\label{sec: summary}

Recently, the Subaru HSC observations found the microlensing candidates~\cite{Sugiyama:2026kpv}.
If these events are caused by PBHs, their mass and fraction to dark matter are estimated as $M_\mathrm{PBH} \sim 10^{-7}\text{--}10^{-6}M_\odot$ and $f_{\rm PBH}=\mathcal{O}(10^{-1})$, respectively. 
To produce such a PBH population, the primordial curvature perturbations should be enhanced on the corresponding small scales.

The axionlike curvaton models can realize the enhancement of the curvature perturbations on a certain scale.
Due to the inflationary dynamics of a complex scalar field, whose phase component plays the role of the curvaton, the curvature power spectrum acquires a characteristic scale dependence.
Moreover, the curvature perturbations induced by the curvaton generically have non-Gaussianity with the sign and magnitude controlled by the curvaton-to-radiation energy ratio at decay, $r_\mathrm{dec}$. 
Such non-Gaussianity can significantly enhance or suppress PBH formation.
Consequently, even with a fixed PBH mass function, the amplitude of the accompanying scalar-induced gravitational waves can vary by orders of magnitude.

In this paper, we have studied production of PBHs in both type I and Type II axionlike curvaton models and showed that the model can produce PBHs with mass $M_\mathrm{PBH}\sim 10^{-7}$--$10^{-6}\,M_\odot$ and abundance $f_\mathrm{PBH}\sim 10^{-1}$ that explain the microlensing candidates reported by the Subaru HSC observations~\cite{Sugiyama:2026kpv} (Fig.~\ref{fig: fPBH}).
Furthermore, we investigated the scalar-induced gravitational waves in the axionlike curvaton models.
The predicted GW spectra have peaks around $10^{-5}\,\mathrm{Hz}$, and will be detected by future GW detectors (Fig.~\ref{fig: OmegaGW}).
Due to the different signs of non-Gaussianities, the curvature perturbations reproducing the microlensing PBHs are larger for $r_\mathrm{dec} = 3$ than for $r_\mathrm{dec}  = 0.3$.
While the peak shape of the curvature power spectrum is similar for $r_\mathrm{dec} = 3$ and $0.3$, the PBH mass function is sharper for $r_\mathrm{dec} = 3$ than for $r_\mathrm{dec} = 0.3$, which is also due to non-Gaussianity.
When comparing type I and II models for $r_\mathrm{dec} = 0.3$, the shape of the curvature power spectrum is distinct, which is inherited by the power spectrum of gravitational waves.
On the other hand, the peak amplitude of the curvature perturbations and the shape of the PBH mass function are similar in both types.

If the baryon asymmetry or dark matter is generated before the curvaton decay, the curvaton fluctuations induce the baryon or CDM isocurvature perturbations~\cite{Lyth:2002my}.
This effect is significant for $r_\mathrm{dec} > 1$.
In this case, the power spectrum of the isocurvature perturbations is given by $\mathcal{P}_S \simeq 9 \mathcal{P}_\zeta$ on the scales where the total curvature perturbation is dominated by the curvaton contribution.
The relevant constraint on the baryon isocurvature perturbations comes from the big bang nucleosynthesis as $\mathcal{P}_{S_b} \lesssim 0.016$ on $k \lesssim 4 \times 10^8 \mathrm{Mpc}^{-1}$~\cite{Inomata:2018htm}.
Depending on the curvature power spectrum, we require that the baryon asymmetry be generated after the curvaton decay.

Finally, we comment on the limitations of our analysis.
In the type II model, we neglected the curvature perturbations on small scales with $k > k_\mathrm{dec}$. 
Although such modes are suppressed compared to those around the peak wavenumber, they can affect the spectrum of the gravitational waves in high-frequency regions, which may be crucial for testing the model.
Moreover, if the curvaton suddenly decays in the case of $r_\mathrm{dec} > 1$, the background universe undergoes a sudden transition from the matter-dominated era to the radiation-dominated era.
Such a transition could lead to the enhancement of the gravitational waves induced by adiabatic curvature perturbations via the so-called poltergeist mechanism~\cite{Inomata:2019ivs}.
The presence of the isocurvature perturbations could further modify the poltergeist mechanism, which is beyond the scope of this study.

\begin{acknowledgments}
We are grateful to Masahiro Takada and Sunao Sugiyama for helpful discussions. 
This work was supported in part by JSPS KAKENHI Grant Numbers d 25KJ1164 (K.K), 25K07297 (M.K.), 23KJ0088 (K.M.), and 25KJ1030 (S.N.), and JST SPRING Grant Number JPMJSP2108 (K.K. and S.N.).
M.K. was supported by World Premier International Research Center Initiative (WPI), MEXT, Japan.
\end{acknowledgments}

\appendix

\section{Relations of wavenumbers before and after inflation}
\label{app: wavenumber}

Here, we derive the relation between wavenumbers before and after inflation.
First, we derive the relation~\eqref{eq: k relation} following Ref.~\cite{Inomata:2023drn}.
We can relate the relative scales of a wavenumber before and after inflation by
\begin{align}
    \frac{k}{k_\mathrm{eq}}
    =
    \frac{k}{\mathcal{H}_\mathrm{end}}
    \frac{\mathcal{H}_\mathrm{end}}{\mathcal{H}_\mathrm{R}}
    \frac{\mathcal{H}_\mathrm{R}}{\mathcal{H}_\mathrm{dec}}
    \frac{\mathcal{H}_\mathrm{dec}}{\mathcal{H}_\mathrm{eq}}
    \ .
\end{align}
Since $\rho^\mathrm{tot} \propto H^2 \propto a^{-3}$ during the MD era, we obtain
\begin{align}
    \frac{\mathcal{H}_\mathrm{end}}{\mathcal{H}_\mathrm{R}}
    =
    \left( \frac{\rho^\mathrm{tot}_\mathrm{end}}{\rho^\mathrm{tot}_\mathrm{R}} \right)^{1/6}    
    =
    \left( \frac{\rho^\mathrm{tot}_\mathrm{end}}{\frac{\pi^2}{30} g_\mathrm{R} T_\mathrm{R}^4} \right)^{1/6}    
    \ .
\end{align}
From the entropy conservation of radiation, we obtain
\begin{align}
    \frac{\mathcal{H}_\mathrm{R}}{\mathcal{H}_\mathrm{dec}}
    &=
    \left( \frac{g^s_\mathrm{dec,bef} T_\mathrm{dec,bef}^3}{g^s_\mathrm{R} T_\mathrm{R}^3} \right)^{1/3}
    \left( \frac{g_\mathrm{R} T_\mathrm{R}^4}{(1+r_\mathrm{dec})g_\mathrm{dec,bef} T_\mathrm{dec,bef}^4} \right)^{1/2}
    \nonumber \\
    &=
    \left( \frac{g^s_\mathrm{dec,bef}}{g^s_\mathrm{R}} \right)^{1/3}
    \left( \frac{g_\mathrm{R}}{(1+r_\mathrm{dec})g_\mathrm{dec,bef}} \right)^{1/4}
    \left( \frac{g_\mathrm{R} T_\mathrm{R}^4}{(1+r_\mathrm{dec})g_\mathrm{dec,bef} T_\mathrm{dec,bef}^4} \right)^{1/4}
    \ .
\end{align}
Similarly, we also obtain
\begin{align}
    \frac{\mathcal{H}_\mathrm{dec}}{\mathcal{H}_\mathrm{eq}}
    &=
    \left( \frac{g^s_\mathrm{eq} T_\mathrm{eq}^3}{g^s_\mathrm{dec,aft} T_\mathrm{dec,aft}^3} \right)^{1/3}
    \left( \frac{g_\mathrm{dec,aft} T_\mathrm{dec,aft}^4}{2 g_\mathrm{eq} T_\mathrm{eq}^4} \right)^{1/2}
    \nonumber \\
    &=
    \left( \frac{g^s_\mathrm{eq}}{g^s_\mathrm{dec,aft}} \right)^{1/3}
    \left( \frac{g_\mathrm{dec,aft}}{2 g_\mathrm{eq}} \right)^{1/4}
    \left( \frac{g_\mathrm{dec,aft} T_\mathrm{dec,aft}^4}{2 g_\mathrm{eq} T_\mathrm{eq}^4} \right)^{1/4}
    \ .
    \label{eq: Hdec to Heq}
\end{align}
Since $(1+r_\mathrm{dec})g_\mathrm{dec,bef} T_\mathrm{dec,bef}^4 = g_\mathrm{dec,aft} T_\mathrm{dec,aft}^4$, we finally obtain
\begin{align}
    \frac{k}{k_\mathrm{eq}}
    =&
    \frac{k}{\mathcal{H}_\mathrm{end}}
    \left( \frac{\rho^\mathrm{tot}_\mathrm{end}}{\frac{\pi^2}{30} g_\mathrm{R} T_\mathrm{R}^4} \right)^{1/6}   
    \left( \frac{g^s_\mathrm{dec,bef}}{g^s_\mathrm{R}} \right)^{1/3}
    \left( \frac{g_\mathrm{R}}{(1+r_\mathrm{dec})g_\mathrm{dec,bef}} \right)^{1/4}
    \left( \frac{g_\mathrm{R} T_\mathrm{R}^4}{(1+r_\mathrm{dec})g_\mathrm{dec,bef} T_\mathrm{dec,bef}^4} \right)^{1/4}
    \nonumber \\
    & \times 
    \left( \frac{g^s_\mathrm{eq}}{g^s_\mathrm{dec,aft}} \right)^{1/3}
    \left( \frac{g_\mathrm{dec,aft}}{2 g_\mathrm{eq}} \right)^{1/4}
    \left( \frac{g_\mathrm{dec,aft} T_\mathrm{dec,aft}^4}{2 g_\mathrm{eq} T_\mathrm{eq}^4} \right)^{1/4}
    \nonumber \\
    =&
    \frac{k}{\mathcal{H}_\mathrm{end}}
    \left( \frac{\rho^\mathrm{tot}_\mathrm{end}}{\frac{\pi^2}{30} g_\mathrm{R} T_\mathrm{R}^4} \right)^{1/6}   
    \left( \frac{g^s_\mathrm{eq} g^s_\mathrm{dec,bef}}{g^s_\mathrm{R} g^s_\mathrm{dec,aft}} \right)^{1/3}
    \left( \frac{g_\mathrm{R} g_\mathrm{dec,aft}}{2(1+r_\mathrm{dec})g_\mathrm{dec,bef} g_\mathrm{eq}} \right)^{1/4}
    \left( \frac{g_\mathrm{R} T_\mathrm{R}^4}{2 g_\mathrm{eq} T_\mathrm{eq}^4} \right)^{1/4}
    \nonumber \\
    =&
    1.3 \times 10^{24} 
    (1 + r_\mathrm{dec})^{-1/4}
    \left( \frac{\rho^\mathrm{tot}_\mathrm{end}}{3 \times 10^{-11} M_\mathrm{Pl}^4} \right)^{1/6}
    \left( \frac{T_\mathrm{R}}{10^{14}\,\mathrm{GeV}} \right)^{1/3}
    \left( \frac{g_\mathrm{R}}{g^s_\mathrm{R} } \right)^{1/3}
    \left( \frac{g^s_\mathrm{dec,bef}}{g^s_\mathrm{dec,aft}} \right)^{1/3}
    \left( \frac{g_\mathrm{dec,aft}}{g_\mathrm{dec,bef}} \right)^{1/4}
    \frac{k}{\mathcal{H}_\mathrm{end}}
    \ ,
\end{align}
where we used $g_\mathrm{eq} = 3.36$, $g^s_\mathrm{eq} = 3.91$, and $T_\mathrm{eq} = 8.0 \times 10^{-10}$\,GeV.
Note that this relation can be used for $r_\mathrm{dec}>1$ and $r_\mathrm{dec} \leq 1$.

Next, we derive the relation for $k_\mathrm{dec} \equiv \mathcal{H}_\mathrm{dec}$.
Since we assume that the curvaton starts to oscillate while the inflaton oscillation dominates the universe (see Fig.~\ref{fig: Energy}), at the onset of oscillations, the energy density of the curvaton and radiation is given by
\begin{align}
    \rho_\sigma = \frac{1}{2} m_\sigma^2 v^2 \theta_i^2 
    \ , \quad 
    \rho_\mathrm{rad} = 3 M_\mathrm{Pl}^2 m_\sigma^2
    \ .
\end{align}
The ratio of these quantities, $r$, is conserved until the inflaton decay.
After that, $r$ evolves as 
\begin{align}
    r(T)
    =
    \frac{v^2 \theta_i^2}{6 M_\mathrm{Pl}^2}
    \frac{g^s T^3}{g^s_\mathrm{R} T_\mathrm{R}^3}
    \frac{g_\mathrm{R} T_\mathrm{R}^4}{g T^4}
    =
    \frac{v^2 \theta_i^2}{6 M_\mathrm{Pl}^2}
    \frac{g^s}{g^s_\mathrm{R}}
    \frac{g_\mathrm{R}}{g}
    \frac{T_\mathrm{R}}{T}
    \ ,
\end{align}
until the curvaton decay.
From this relation, we obtain
\begin{align}
    T_\mathrm{dec,aft}
    =
    \left( \frac{g_\mathrm{dec,bef}}{g_\mathrm{dec,aft}} \right)^{1/4}
    (1+r_\mathrm{dec})^{1/4}
    T_\mathrm{dec,bef}
    =
    \left( \frac{g_\mathrm{dec,bef}}{g_\mathrm{dec,aft}} \right)^{1/4}
    (1+r_\mathrm{dec})^{1/4}
    \frac{v^2 \theta_i^2}{6 M_\mathrm{Pl}^2}
    \frac{g^s_\mathrm{dec,bef}}{g^s_\mathrm{R}}
    \frac{g_\mathrm{R}}{g_\mathrm{dec,bef}}
    \frac{T_\mathrm{R}}{r_\mathrm{dec}}
    \ .
\end{align}
By substituting this relation to Eq.~\eqref{eq: Hdec to Heq}, we find 
\begin{align}
    \frac{k_\mathrm{dec}}{k_\mathrm{eq}}
    =
    \left( \frac{g^s_\mathrm{eq}}{g^s_\mathrm{dec,aft}} \right)^{1/3}
    \left( \frac{g_\mathrm{dec,aft}}{2 g_\mathrm{eq}} \right)^{1/2}
    \left( \frac{g_\mathrm{dec,bef}}{g_\mathrm{dec,aft}} \right)^{1/4}
    \frac{g^s_\mathrm{dec,bef}}{g^s_\mathrm{R}}
    \frac{g_\mathrm{R}}{g_\mathrm{dec,bef}}
    \frac{(1+r_\mathrm{dec})^{1/4}}{r_\mathrm{dec}}
    \frac{v^2 \theta_i^2}{6 M_\mathrm{Pl}^2}
    \frac{T_\mathrm{R}}{T_\mathrm{eq}}
    \ .
\end{align}

\section{Type I axionlike curvaton model}
\label{app: type_I_model}

In this appendix, we introduce the Type I axionlike curvaton model in the framework of supersymmetry. 
Let us consider the superpotential given by
\begin{equation}
    W = h S\left(\Phi\bar{\Phi} -\frac{v^2}{2}\right),
\end{equation}
where $\Phi$, $\bar{\Phi}$, and $S$ are chiral superfields with charges $+1$, $-1$, and $0$, respectively, under a global $U(1)$ symmetry, and $h$ is a coupling constant.
This superpotential leads to the scalar potential,
\begin{equation}
    V = h^2\left|\Phi\bar{\Phi}-\frac{v^2}{2}\right|^2 +h^2|S|^2(|\Phi|^2+\bar{\Phi}|^2),
\end{equation}
where the scalar component of the superfield is denoted by the same symbol. 
The potential is flat along the following direction:
\begin{equation}
    \Phi\bar{\Phi} = \frac{v^2}{2}, \qquad S=0.
\end{equation}
This flat direction is lifted by the supergravity effect, which provides the Hubble-induced masses as 
\begin{equation}
    V_H = c H^2 |\Phi|^2+ \bar{c}H^2|\bar{\Phi}|^2
    +c_S H^2|S|^2,
\end{equation}
where $c$, $\bar{c}$, and $c_S$ are constants.
Owing to the Hubble-induced terms, the flat direction has a minimum $|\Phi|\sim |\bar{\Phi}|\sim v$.

In the axionlike curvaton model, we assume $|\Phi| \gg |\bar{\Phi}|$ at the beginning of inflation.
This allows us to focus only on the dynamics of $\Phi$ with the potential $cH^2|\Phi|^2$ at $|\Phi|\gg v$.
The $\Phi$ field is decomposed into the radial and phase directions as
\begin{equation}
    \Phi = \frac{\varphi_0}{\sqrt{2}}\exp\left(i\frac{\sigma}{v}\right).
\end{equation}
Taking into account that $\phi$ is stabilized at $\varphi_0 \sim v$, the effective potential for $\varphi_0$ is given by
\begin{equation}
    V(\varphi_0) \simeq \frac{c}{2}H^2 (\varphi_0-v)^2.
\end{equation}

\section{PBH mass}
\label{app: PBH mass}

Here, we derive the formulae related to PBHs used in the main text.
First, we derive the relation between the PBH mass and the corresponding wavenumber following Ref.~\cite{Inomata:2018cht}.
From the relations of 
\begin{align}
    \frac{k_\mathrm{PBH} }{a_\mathrm{form}}
    &=
    H_\mathrm{form}
    \ ,
    \\
    \rho_\mathrm{tot}^\mathrm{form}
    &=
    3 M_\mathrm{Pl}^2 H_\mathrm{form}^2
    =
    \frac{\pi^2}{30} g_\mathrm{form} T_\mathrm{form}^4
    \ ,
    \\
    a_\mathrm{form}
    &=
    \left( \frac{g^s_0}{g^s_\mathrm{form}} \right)^{1/3}
    \frac{T_0}{T_\mathrm{form}}
    \ ,
\end{align}
we obtain 
\begin{align}
    M_\mathrm{PBH}(k_\mathrm{PBH})
    &=
    \gamma \rho^\mathrm{tot}_\mathrm{form} \frac{4\pi H_\mathrm{form}^{-3}}{3} 
    =
    4 \pi \gamma M_\mathrm{Pl}^2 
    H_\mathrm{form}^{-1}
    \nonumber \\
    &\simeq 
    M_\odot 
    \left( \frac{\gamma}{0.2} \right)
    \left( \frac{g_\mathrm{form}}{10.75} \right)^{1/2}
    \left( \frac{g^s_\mathrm{form}}{10.75} \right)^{-2/3}
    \left( \frac{k_\mathrm{{PBH}}}{2.0 \times 10^{-6}\,\mathrm{Mpc}^{-1}} \right)^{-2} 
    \ ,
\end{align}
where we used $T_0 = 2.73$\,K and $g^s_0 = 3.91$.

Next, we derive the relation of $f_\mathrm{PBH}$ and $\beta$:
\begin{align}
    f_\mathrm{PBH}
    &=
    \left. 
        \frac{1}{{\rho_\mathrm{m}}}
        \frac{\mathrm{d} \rho_\mathrm{PBH}}{\mathrm{d} \ln M_\mathrm{PBH}}
    \right|_\mathrm{eq}
    \frac{\Omega_\mathrm{m}}{\Omega_\mathrm{DM}}
    \nonumber \\
    &=
    \gamma \beta(M_\mathrm{PBH})
    \frac{g^s_\mathrm{eq} T_\mathrm{eq}^3}{g^s_\mathrm{form} T_\mathrm{form}^3}
    \frac{g_\mathrm{form} T_\mathrm{form}^4}{g_\mathrm{eq} T_\mathrm{eq}^4}
    \frac{\Omega_\mathrm{m}}{\Omega_\mathrm{DM}}
    \nonumber \\
    &=
    \gamma \beta(M_\mathrm{PBH})
    \frac{g^s_\mathrm{eq}}{g^s_\mathrm{form}}
    \frac{g_\mathrm{form}}{g_\mathrm{eq}}
    \frac{T_\mathrm{form}}{T_\mathrm{eq}}
    \frac{\Omega_\mathrm{m}}{\Omega_\mathrm{DM}}
    \nonumber \\
    &=
    \frac{\beta(M_\mathrm{PBH})}{1.7 \times 10^{-8}}
    \left( \frac{\gamma}{0.2} \right)^{3/2}
    \left( \frac{g_\mathrm{form}}{10.75} \right)^{3/4}
    \left( \frac{g^s_\mathrm{form}}{10.75} \right)^{-1}
    \left( \frac{M_\mathrm{PBH}}{M_\odot} \right)^{-1/2}
    \ ,
\end{align}
where we used $\Omega_\mathrm{DM} h^2 = 0.12$ and $\Omega_\mathrm{m} h^2 = 0.142$.

\bibliographystyle{JHEP}
\bibliography{Ref}

\end{document}